\documentclass[a4paper,11pt]{article}
\usepackage{nfraconf,graphicx,epsf}

\begin{document}
\baselineskip=13pt

\title{Prospects for solving basic questions in galaxy
evolution with a next generation radio telescope}

\author{ F.H. Briggs}
\address{Kapteyn Astronomical Institute\\
P.O. Box 800\\
9700 AV Groningen\\
The Netherlands\\
E-mail: fbriggs@astro.rug.nl}

\maketitle

\abstract{A Square Kilometer Array radio telescope will detect
tens of thousands of galaxies per square degree in the 21cm emission
line of neutral hydrogen. The telescope will be sensitive to ordinary galaxy
populations at redshifts $z>3$ when the mass density in neutral
gas greatly exceeded the mass in luminous stars.  Spectroscopy in
the HI line will 
trace the kinematics of evolving galactic potentials, as well 
as monitoring  consumption of the fuel for star formation.
The OH megamaser population may be a significant indicator
of merger and interaction rates for galaxies over cosmic time.
Spectroscopy of the redshifted 
 HI and OH lines provides a major incentive for designing
a large radio telescope with continuous frequency
coverage below 2~GHz. 
}
 
\section{Introduction}
 
A number of indicators point to the period from $z\approx 3$ to
$z\approx 1$ as the time when galaxies
assembled.  This period of some 3 billion years witnessed
a maximum in the rate at which stars have formed and a
peak in number of quasars and powerful radio galaxies.  
The luminous star forming galaxies that
trace the rise and fall of star formation through this epoch
represent a tiny fraction
of the protogalactic systems that exist at this time,
since neutral hydrogen clouds are known to have been  abundant,
amounting to far more mass than was then in stars.
This leaves the bulk of the baryons destined to join
galaxies below the threshold for
viewing by today's telescopes, and it means that our perception of
this important epoch of history lacks a clear observational
picture of the sequence and timing of events that has occurred
in the coalescence of mass to form galactic potential wells and
their  present contents of stars, gas and dust.  

Theoretical simulations, which are successfully tuned to produce the $z=0$
large scale structure starting from initial conditions that are consistent
with the level of fluctuation in the microwave background, predict
mass distribution functions for the protogalaxies through this
period of galaxy formation.
These simulations, which are based on the gravitational collapse
of a Cold Dark Matter dominated Universe, demonstrate a hierarchical
clustering that leads to the desired $z\approx 0$ large-scale behavior and
shows galaxies forming by the dissipationless
merging of low mass dark matter mini-halos halos
and the subsequent accretion of condensing, dissipational gas
(cf. White \& Frenk 1991, Kauffman 1996, Ma et al 1997).

The postponement of the assembly of massive galaxies in the
CDM models  is somewhat at odds with observations showing
powerful radio galaxies at redshift above $z=5$ (van Breugel et al 1999)
and evidence for large gaseous disks in well-formed potentials 
at $z\approx 2.5$ (Prochaska \& Wolfe 1999).  A crucial test of
the formation ideas will be to measure the sizes of galaxies and
their total dynamical masses as a function of redshift in order
to define time sequence of galaxy formation.

The next 
section reviews the considerable indirect evidence that bears
on this epoch, leading to the conclusion that a large radio telescope,
capable of sensing the abundant cool gas confined to the
evolving potential wells, will clarify the history of galaxies
through this critical period.  Subsequent sections address the
specific observational tests.

\section{Evolution of global properties $z=4$ to $z=0$}
 
\begin{figure}
  \begin{center}
    \includegraphics[width=15cm ]{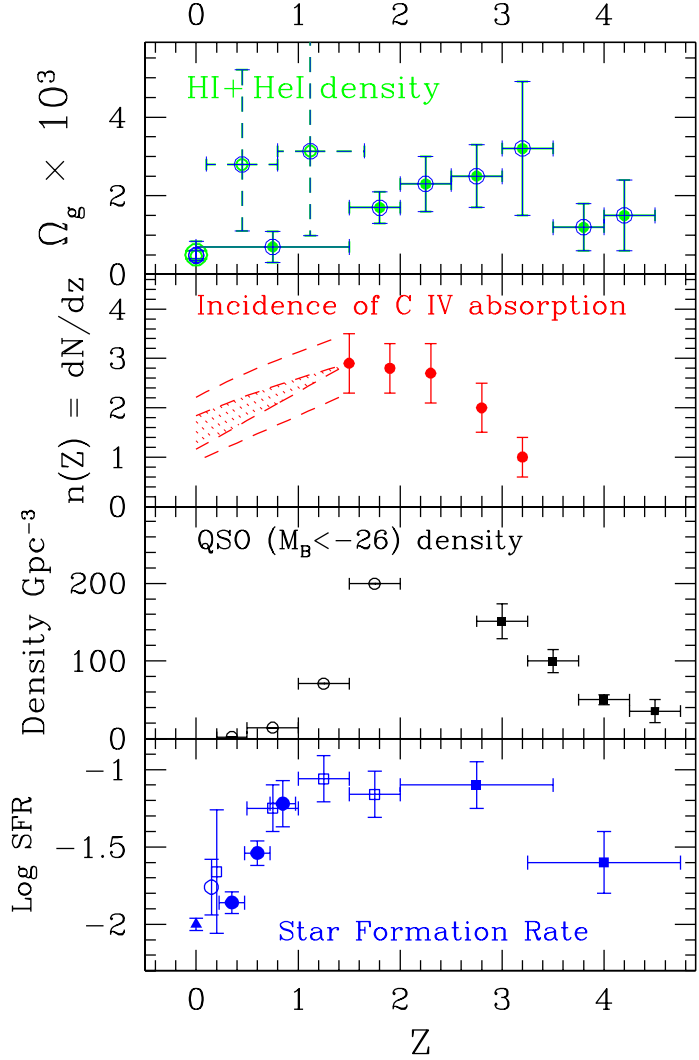}
    \caption{\small Cosmological density of neutral gas, incidence 
of CIV absorption, comoving density of luminous QSOs, and mean
star formation rate as a function of redshift. 
{\it Top panel.}  Mean cosmological density of neutral gas,
$\Omega_g$, normalized to the critical density (Storrie-Lombardi et al
1996;   Lanzetta et al 1995, Zwaan et al 1997($z=0$), 
Turnshek 1998: dashed error bars)
{\it Upper middle panel.} Number of CIV metal-line absorption systems per unit redshift, $n(z)$ (Steidel 1990).  
Hatched areas indicate the range ($0<q_o<1/2$) for unevolving
cross sections since $z=1.5$, beyond which redshift CIV can be
measured with ground-based telescopes. 
{\it Lower middle panel.} Comoving density of optically selected QSOs:
filled squares from Schmidt et al 1994; open circles from Hewitt et al 1993).
$H_o = 50 $km~s$^{-1}$Mpc$^{-1}$, $q_o =1/2$
{\it Bottom panel.} Comoving star formation rate density 
$M_{\odot}$yr$^{-1}$Mpc$^{-3}$ from Madau (1998) and references therein.
}
    \label{hi_civ_q_sf.fig}
  \end{center}
\end{figure} 

A number of indicators trace global properties of the Universe
through the epoch of most vigorous assembly of galaxies.
Fig. 1 shows four of these plotted as a function of
redshift. (1)~The star formation rate SFR density computed for
color-selected,  star forming galaxies (cf Madau et al 1996, Calzetti
and Heckman 1999) shows a steep rise with redshift to $z\approx 1$ with
a modest decline at higher redshifts. More recent observational
evidence (Steidel et al 1999) favors a flat SFR density
above $z=1$ to at least $z=4$, once corrections have been made for
extinction. (2)~The comoving space density of
luminous optically selected quasars reaches a maximum at $z\approx 2$,
implying that mass is being redistributed within the evolving 
galaxies to efficiently feed nuclear activity. 
(3)~Studies of absorption lines against the ultraviolet continua of bright
high redshift quasars provide probes of the comoving density of neutral
gas, through studies of the damped Lyman-$\alpha$ DLa line of HI (Wolfe
et al 1986), as
well as (4)~the ionized galaxy halo gas that is sensed in CIV (Steidel 1990).
Quasar absorption lines are especially relevant for the studies of
normal galaxies, since they are not biased toward the luminous objects
at the peak of the luminosity function for a chosen redshift, but 
rather the quasar absorption lines provide a democratic selection of the
common, gas rich objects that represent the 
less rapidly evolving protogalaxies
that are contain the bulk of the baryons destined to
eventually be locked into galaxies at $z=0$.

The DLa lines are especially relevant to the discussion
here, since the quantities of cool neutral gas sensed in these
high redshift absorbers exceeds the local HI comoving density $\Omega_g(z=0)$
by a factor of at least five (Wolfe et al 1995, Lanzetta 1995), and
the HI is a viable target for radio studies in the 21cm line of HI.
The DLa class comprises neutral gas layers with H$^{0}$ column densities
above $2{\times}10^{20}$cm$^{-2}$, while the MgII selection chooses
column densities of H$^{0}$ down to levels  ${\sim}10^{18}$cm$^{-2}$.

The HI content contained in the DLa population of absorbing cloud suffers
uncertainty in the high redshift regime. The same clouds that
absorb in the Lyman-$\alpha$ line, also show absorption by common metals
that originated in a prior generation of stars. The same generation of
stars will have produced dust, and the resulting extinction may
cause lines of sight with DLa absorbers to be selected against when
samples of high $z$ quasars are composed for spectroscopic study
(Pei et al 1999). A detailed appraisal of the extinction leads Pei
et al (1999) to conclude that the DLa statistics may underestimate the
neutral gas density at $z\approx 2.5$ by a factor of two to three, for
a total factor of 10 to 15 above the present density $\Omega_g(z=0)$.

A large uncertainty also remains in the low $0<z<1.7$ 
DLa measures of $\Omega_g(z)$
since the Lyman-$\alpha$ line is not shifted in the optical window
observable with ground based telescopes until $z>1.7$.  This is a
period of galaxy evolution when the SFR subsided, perhaps in response
to depleted supplies of neutral gas from which to form stars. During this
period, star formation may have changed from being fueled with largely   
primordial material to star formation relying on reprocessing an interstellar
medium of built from stellar mass loss (Kennicutt et al 1994) of
earlier generations of relatively long lived stars. An understanding
of this transition is important in predicting the gas-consumption time
(Roberts 1963) and the duration of star formation into the future.
These uncertainties could be eliminated by using a SKA class telescope
to observe the HI emission directly in a survey to redshift $z=1.5$
to perform the kinds of HI census that is being made for the local
Universe now by surveys conducted with telescopes 
such as Arecibo (Zwaan et al 1997) and
Parkes (Staveley-Smith et al 1996).

Currently operational telescopes map
 the detailed gas kinematics of nearby spiral galaxies   in 
the 21cm emission line,
and the analyses show that  galaxies rich in neutral gas
have well ordered rotation of a cool disk, much like the disk
of the Milky Way. The dynamical analysis of the velocity fields of
the quiescent rotating disks has been a important tool in the
measurement of the total mass in galaxies and has served to specify the
presence and distribution of dark matter in galaxies
(cf van Albada et al 1985,
Broeils 1992). In this application, the HI is a highly diagnostic tracer
of the galactic potential.  Even the crude resolution obtained with
single-dish telescopes yields a ``mass indicator'' by measurement
of the profile width.  The settled and quiescent nature of
neutral gas layers makes them a more reliable tracer of 
gravitational potentials than emission line gas in HII regions
around star forming regions, where stellar winds and expanding
shock fronts compete with gravitation in determining the 
gas kinematics.

The apparent absence of isolated neutral gas clouds without an associated
galaxy in the nearby Universe (Zwaan et al 1997, Spitzak and
Schneider 1999) suggests that the neutral gas relies on
a confining potential to maintain the gas at sufficient density that the
it can remain neutral in the face of an ionizing background
radiation field. Perhaps under the conditions of ``adequate'' confinement
for preservation of neutrality, it is difficult to avoid the
instabilities that lead to cooling, collapse and star formation. 
In such a picture, the shallower potential wells of the lower mass
dwarf galaxies would be the places where the HI is more gently 
confined and evolutionary processes would generally proceed 
more slowly. Indeed, this is the domain inhabited by the dimmest
of the gas-rich LSB galaxies.  The expectation is that the
DLa clouds at high redshift must also be confined in order to
remain neutral, and they too will be tracers of their confining
potentials.

In many respects, the neutral clouds giving rise to the
the DLa absorbers have similar properties to the interstellar media of gas
rich galaxies. They typically have a mix of cold clouds and a thinner,
turbulent medium (Briggs \& Wolfe 1983, Lanzetta \& Bowen 1992,
Prochaska \& Wolfe 1997) whose physical conditions vary
from mildly ionized to the more highly
ionized ``halo'' gas characterized by the  CIV absorption lines
and represented by well studied lines of sight through the
Milky Way halo. Metal abundances in the DLa clouds show considerable
variance around the expected trend of enrichment over time (Pettini
et al 1997), and there is now evidence that the DLa clouds are a distinct
population (Pettini et al 1999)
apart from the active star forming galaxies found through
color selection (Steidel et al 1999). The onset of the CIV absorption
(see Fig.~\ref{hi_civ_q_sf.fig}) is another symptom of wide-spread
star formation, as metals are produced and redistributed in the ISM and halos
(Steidel 1990, Sembach et al 1999). 

\section{Gas-rich clouds vs. active star forming galaxies}

\begin{figure}
\hsize .927in
\vglue  .3in
\hglue 11cm\epsfxsize=\hsize\epsffile{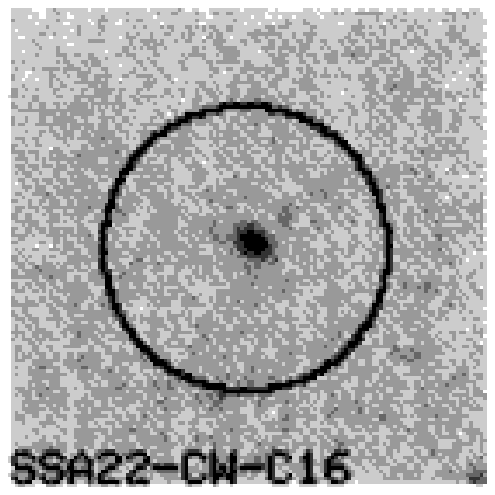}
\hsize 1.5in
\vglue 1.6in
\hsize 4.05in
\vglue -3.2in
\hglue +1.2cm\epsfxsize=\hsize\epsffile{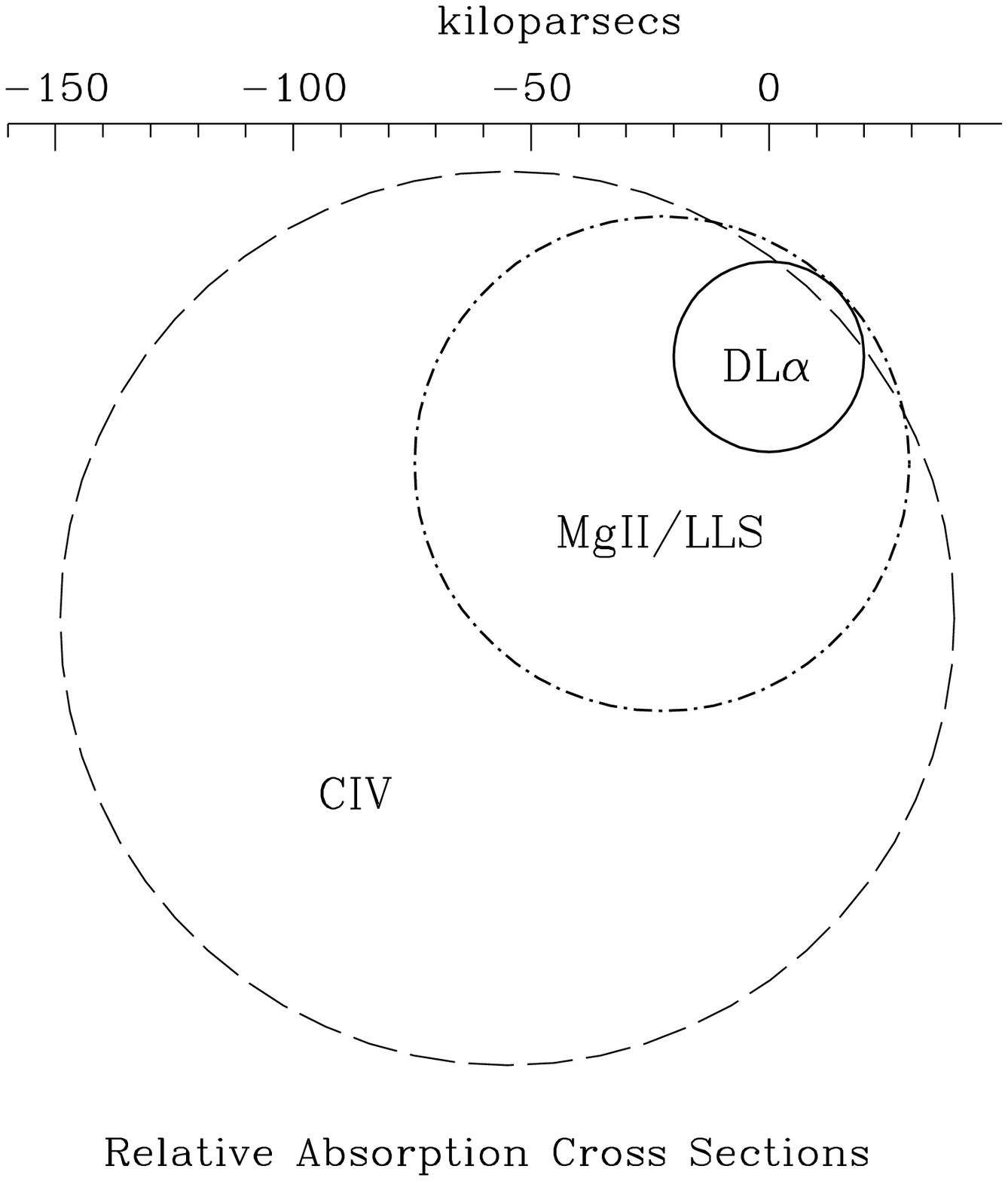}
\hsize 14.5cm
\caption{ Comparison of quasar absorption-line
cross sections for CIV, MgII-Lyman Limit, and damped
Lyman$-\alpha$ lines with the physical size of the optical emission
from  a color-selected galaxy at $z\approx 3$ {\it top right} 
(Giavalisco et al 1996a).   The
$z\approx 3$ galaxy is centered in a 5$''$ diameter circle that
subtends 37.5 kpc ($H_o=75$~km~s$^{-1}$Mpc$^{-1}$, $\Omega=0.2$).
}
\label{crosssection.fig}
\end{figure}

An interesting comparison can be made between the observed sizes of the
high-$z$ star forming galaxies (Giavalisco et al 1996a) and the
interception cross-sections for uv absorption by different ions
(cf. Steidel 1993).  The Lyman-break color-selection technique
for identifying the star forming galaxies produces candidates 
with a density on the sky of ${\sim}1$~arcmin$^{-2}$ for objects
with redshifts predominantly in the range $2.6\leq z\leq 3.4$
(Steidel et al 1998).
The comoving density of $L\geq L_*$ galaxy ``sites,'' computed for this redshift
range, amounts to ${\sim}2$~arcmin$^{-2}$ (for a cosmological model
with $\Omega_o=0.2$). Fig.~\ref{crosssection.fig} shows 
the  cross section for absorption lines  that every
$L_*$ galaxy site would necessarily present, if ordinary galaxies
are to explain the observed incidence of absorption lines.  
Thus, absorption line statistics indicate $\sim$2 times
the absorption cross sections shown in Fig.~\ref{crosssection.fig}
for every Lyman-break galaxy. At low and intermediate redshifts ($z<1$),
the association of the metal line absorbers with the outer regions
of galaxies has been well established for the MgII class of 
absorber by observation of galaxies close to the lines of sight
to the background quasars (Le Brun et al 1993). The CIV selected systems  are 
consistent with galaxy halo cloud properties (Petijean \& Bergeron
1994, Guillemin \& Bergeron 1997), 
and ionized high velocity cloud (HVC) analogs for the 
CIV cloud population exist in the halo of the Milky Way
(Sembach et al 1999). A few of the
DLa absorbers are  also identified with galaxies at intermediate redshift
(Steidel et al 1994, Steidel et al 1995). The optical identification
of the DLa systems has gone more slowly than for the MgII, for
example, because DLa absorbers are rarer (as indicated by the
relative cross sections), and the majority of the surveys for
DLa systems have been conducted with ground-based telescopes, which
find $z>1.7$ systems that are difficult to associate with
galaxies. Curiously, some of the studies of the lowest redshift DLa absorbers
have failed to provide any optical identification to sensitive
limits (Rao \& Turnshek 1998).

The overall implication of a comparison between the physical 
sizes of the Lyman-break galaxies and the absorption-line 
cross sections for the
high-$z$ Universe is that there is a substantial population
of metal-enriched gaseous objects,
possibly accompanied by a tiny pocket of
stellar emission, that can well go unnoticed
in deep optical images. The nature of these invisible absorbers remains
a puzzle.  Do the basic skeletons of today's $L^*$ galaxies
exist already at redshifts greater than 3 as partially filled
gravitational potential wells of dark matter -- each well binding
a star forming nuclear region, a large disk of neutral gas,
 and an extended halo structure of
ionized, metal-enriched gas? Or,
are the statistical cross sections for absorption plotted
in Fig.~\ref{crosssection.fig} actually the integrals of many much
smaller absorption cross sections created by much
less strongly bound, small clouds that 
coalesce steadily since $z\approx 5$ to form the large galaxies at $z=0$?
In the latter case, the star-forming Lyman Break population would
represent a very tiny fraction of the protogalaxy population. Will all
galaxies pass through such a star-forming phase, or will they be 
accreted to the onto the LB objects? Would the tiny DLa clouds need to
be bound in dark matter mini-halos in order to avoid 
photoionization as has occurred in the intergalactic medium?
Are the DLa clouds clustered together or are they bound to the outskirts of
the LB galaxies?  These are important questions since the
DLa population appears to be a gravitationally confined population
containing enough baryons to produce the stellar matter at $z=0$,
and if these clouds are largely cool and unevolved, their
kinematic and dynamic properties can be
studied directly in no way other than at radio wavelengths.

\section{The big questions}

\begin{figure}
  \begin{center}
    \includegraphics[width=13cm ]{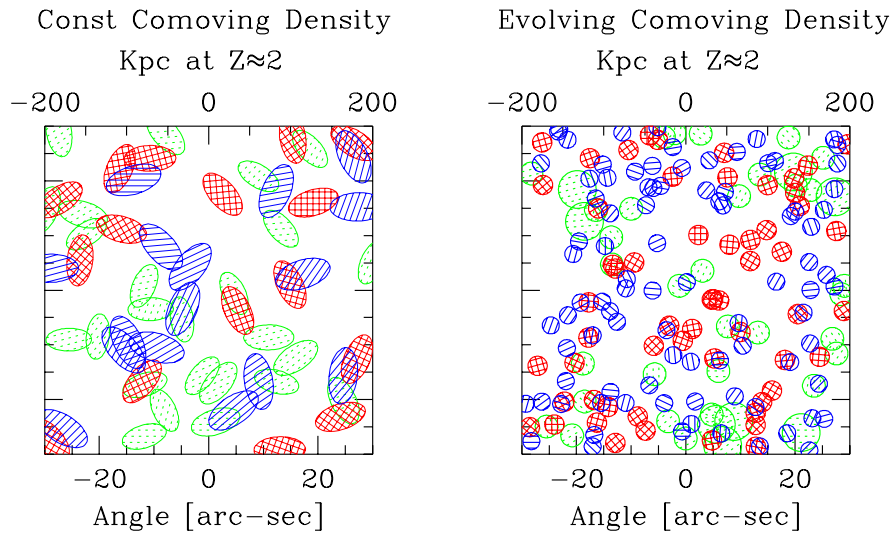}
    \caption{\small Comparison of cross sections presented
in neutral gas. These are views of the coverage of the sky by 
DLa clouds out to $z=4$. {\it Left} Large disk galaxies. {\it Right} Tiny
protogalactic mini-halos.  Galaxies at $z=0$ to 2 are lightly
shaded; $z=2$ to 3 are most heavily shaded; $z=3$ to 4 have medium
shading. Panels are 1 arcmin square, and the top axis gives the
physical scale that applies at $z=2$. 
($H_o=75$~km~s$^{-1}$Mpc$^{-1}$, $\Omega=0.2$)
}
    \label{schematic.fig}
  \end{center}
\end{figure}

The big question is centered on the sequence for construction of
the galaxy population. Do galaxies form in a hierarchical manner
as described by the CDM simulations?  Can merging and accretion be
gentle enough to produce the cool flattened disk galaxies like 
the spiral population at $z\approx 0$?

Fig.~\ref{schematic.fig} is a schematic view of the observational
consequences for the two scenarios -- one with large objects already
in place at high $z$ and one relying on hierarchical merging.
In order to build up the cross section
required by the DLa statistics 
at $z\approx 2.5$, there must be several disks per unit
$\Delta z$ per square arcminute along a randomly chosen direction.
The left panel in Fig.~\ref{schematic.fig} illustrates this
idea with disks whose comoving number density is constant over time
from
$z=4$ to $z=0$
and whose diameters are adjusted  to
match the DLa interception probabilities (Lanzetta et al 1995).
There is an ${\sim}$50\% probability of interception by a DLa
over a redshift path from 0 to 4.
The right panel in Fig.~\ref{schematic.fig} has individual 
spherically shaped clouds that decline in radius toward higher 
redshifts as $r\propto (1+0.7z)^{-3/2}$, while maintaining the
same integral interception cross section per unit redshift as in the
left panel. This requires that there be many more clouds to
build up the required integral cross section.
The metal line systems, such as CIV, have  larger cross section, as shown in 
Fig.~\ref{crosssection.fig}, and a similar diagram plotted for CIV statistics
would provide near complete coverage of the sky, with a significant
probability of multiple CIV absorbers along a single line of sight,
as is observed (Steidel 1990). 

The angular size scales of the DLa absorbers range
over a few arcseconds, which is a good match to the sub-arcsecond
resolution obtained by a radio interferometer array with baselines
of a few kilometers to a few hundred kilometers. As depicted in
Fig.~\ref{crosssection.fig}, these objects are everywhere in the
sky, so a sufficiently sensitive telescope would find them in
surveys of randomly chosen fields.  A further consequence of
the existence of vast numbers of these objects is that they
should frequently be seen in absorption against background 
radio sources. Roughly one half of the radio galaxies and
quasars at $z \geq 4$ will lie behind DLa clouds.

The prospects for making definitive emission
and absorption experiments are discussed in the following
sections. These observations include detection and mapping
of HI emission from individual high redshift $z>3$ galaxies,
as well as statistical studies of the HI content contained in
sub-classes of optically selected objects.  Measuring 21cm 
absorption against background sources might be an effective
way to settle the question of spatial extent and dynamical
mass content of the DLa population; the absorption experiments
could be tackled during the next few years -- with existing
telescope arrays.

A consequence of hierarchical formation models is that galaxies
should undergo repeated merging and accretion events. At $z\approx 0$
the most noticeable merging systems are also bright far infrared
emitters and often are hosts for OH megamasers.  The brightest megamasers
are so luminous that they could be seen throughout the $z<5$ Universe.
This may provide a means to directly measure the galaxy merging rate as
a function of time, without bias due to dust obscuration.

\section{Monitoring the HI content of the Universe}

\begin{figure}
  \begin{center}
    \includegraphics[width=13cm ]{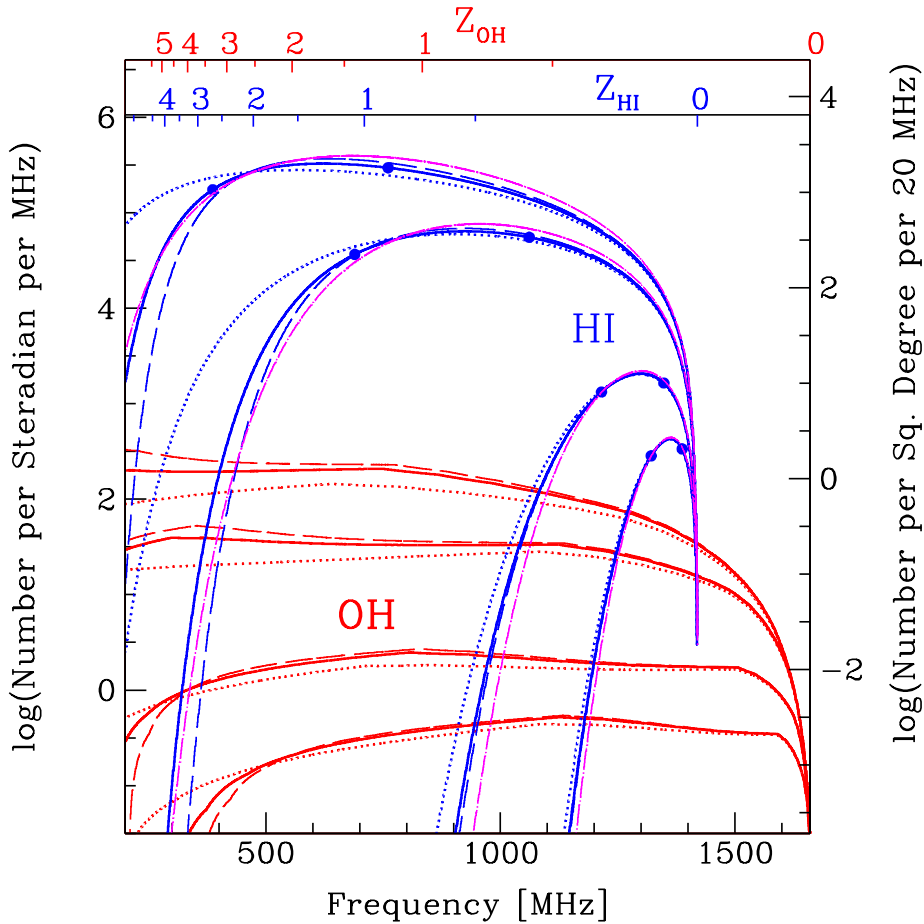}
    \caption{\small Detection rate for HI and OH emission 
from high redshift galaxies, assuming no evolution, as a function
of observed frequency. Top axes indicate corresponding redshifts for the
HI 21cm and OH 18cm emission lines.  Left vertical
axis indicates detections ster$^{-1}$MHz$^{-1}$, and right vertical
axis has detections deg$^{-2}$ per 20 MHz. Detection rates are computed
for detection thresholds of 1000, 200, 5, and 0.75$\mu$Jy.
Cosmological models are: {\it dotted} $\Omega_m= 1$, $\Omega_{\Lambda}= 0$,
{\it solid} $\Omega_m= 0.2$, $\Omega_{\Lambda}= 0$,
{\it dashed} $\Omega_m=0.02$, $\Omega_{\Lambda}= 0$,
{\it dot-dash} $\Omega_m=0.2$, $\Omega_{\Lambda}= 0.8$.
Large dots are drawn for the $\Omega_m= 0.2$, $\Omega_{\Lambda}= 0$
models to indicate the maximum redshifts where $M_{HI}^*$ and $0.1M_{HI}^*$
galaxies could be detected in the HI line.
}
    \label{detecrate.fig}
  \end{center}
\end{figure}

The expectation for detection of HI-rich galaxies at different
redshifts is straightforward to calculate. To provide the
framework for these estimates, Fig.~\ref{detecrate.fig} summarizes
simulations of the
number of detected signals per unit of survey solid angle per
unit of radio frequency. These units are chosen to facilitate
comparison between the sky density of OH megamasers and more
common but less luminous emission from ordinary galaxies in
the 21cm line. The detection rates are computed for a range
of cosmological models.  Computed sensitivity levels assume
signal profile widths of 300 km~s$^{-1}$ with optimally smoothed
spectra. The detection threshold levels are: 1 mJy, 0.2 mJy, 5$\mu$Jy, 
and 0.75$\mu$Jy.  The levels at 5$\mu$Jy and 0.75$\mu$Jy are 7$\sigma$
for 8 and 360 hour integrations respectively with a telescope with the nominal
specifications suggested for SKA (Taylor \& Braun 1999).

The calculation is based on the number density of
galaxies of given HI mass described by the HI mass function of
Zwaan et al (1999), truncated at the low mass end at $10^5$M$_{\odot}$.
This mass function follows a Schechter form, which has power law rise
toward low masses with index $\alpha=1.2$ and a knee at 
$M_{HI}^* = 10^{9.8}$M$_{\odot}$ ($H_o=75$ km~s$^{-1}$~Mpc$^{-1}$)
above which the mass function
cuts off sharply with exponential dependence. 

A SKA Deep Field HI survey would yield ${\sim}10^5$ gas rich
galaxies from a 1 square degree field. The most numerous
detections would probably fall in the redshift range $0.8<z<2$. These
objects would be excellent tracers of large scale structure.

\subsection{Low redshift: $z<1$}

For low redshifts, the Zwaan et al mass function should provide a reasonably
reliable estimate of the detection rate. 
Through the redshift range from $z\approx 1$ to $z=0$, the Deep Field
survey would observe the decline of the HI mass content
of the Universe, as mass is increasingly locked up in stars.
The difference between the star formation rate density (as measured in
optical surveys) and gas consumption rate will specify the
role of stellar mass loss in replenishing the ISM and prolonging
star formation into the future.

\subsection{Direct detection of protogalaxies at $z\geq 2$}

As a framework for discussion, the calculation of Fig.~\ref{detecrate.fig}
has adopted  a constant
comoving density of non-evolving galaxies.  
This cannot be correct
at high redshifts. 
We  know that there is more
neutral mass at $z\approx 2.5$ than at present. Whether this will also
translate into an increased number of detections above what is
specified by these calculations is not clear.
Whether the HI is parceled in large or small masses will be the
deciding factor.
In order to illustrate the difficulty in
measuring small HI masses predicted by hierarchical clustering
scheme, the figure has dots drawn to indicate the highest
redshift at which non-evolved $M_{HI}^*(z=0)$ and 0.1$M_{HI}^*(z=0)$
galaxies would be detected, for the $\Omega_m= 0.2$, $\Omega_{\Lambda}= 0$
cosmology. For a SKA Deep Field requiring 360 hours of integration,
the $M_{HI}^*$ galaxies are detected to redshift $z\approx 2.6$ and
an 0.1$M_{HI}^*$ to $z\approx 0.9$.

The goal of observing reshifts around $z\approx 2.5$ would
be resolve the considerable uncertainty in HI mass distribution.
The large excess of HI that we know exists could simply increase
all galaxy sites by a factor of 10 in mass, implying that an
0.1$M_{HI}^*$ galaxy at present evolved from a system of 
1$M_{HI}^*$ at $z=2.5$, and these would be easily detected,
along with the 10$M_{HI}^*$ around the knee in the mass function.
On the other hand, a hierarchical picture would have increased
numbers, say by a factor 10, of objects with some fraction
of the HI mass now measured in galaxies nearby. Note that as
galaxies merge and stars form, the stellar 
populations form a
sink for neutral baryons -- merging ten  $M_{HI}=10^9$M$_{\odot}$
protogalaxies is likely to produce a luminous $z=0$ galaxy
with only ${\sim}10^9$M$_{\odot}$ or less of HI mass, since 
much of the mass is destined to be consumed in star formation.

Individual protogalactic clumps with $M_{HI}<10^8$M$_{\odot}$
would not be detected by even the long integrations of a
Deep Field survey. However, the coalescing clumps may be
clustered sufficiently to create  detectable signals. 
 
\subsection{A statistical measure of HI in the Lyman Break Population}

The detection of the small, protogalactic HI masses common to 
hierarchical models at redshifts around $z\approx 3$ will
be difficult, even with a SKA class telescope. On the
other hand, considerable progress might be made with a straightforward
statistical method, much sooner than the 
construction of a new radio telescope. Several current generation
aperture synthesis telescopes (the Westerbork Synthesis Radio Telescope
and the Giant Metrewave Radio Telescope in India)
are equipped to observe the 21cm line redshifted to $z\approx 3$.
The field of view of these telescopes can survey several square
degrees of sky in a single integration with sufficient angular
resolution to avoid confusion among the LB galaxies. If an adequate
catalog of LB galaxies could be constructed for such a synthesis field,
with of order $10^4$ identified LB objects 
with celestial coordinates and redshifts,
then the radio signals could be stacked, to obtain a statistical measure
of the HI content of the LB population.  This would allow the
``average HI content per LB galaxy site'' to be determined. 

\begin{figure}
  \begin{center}
    \includegraphics[width=10cm ]{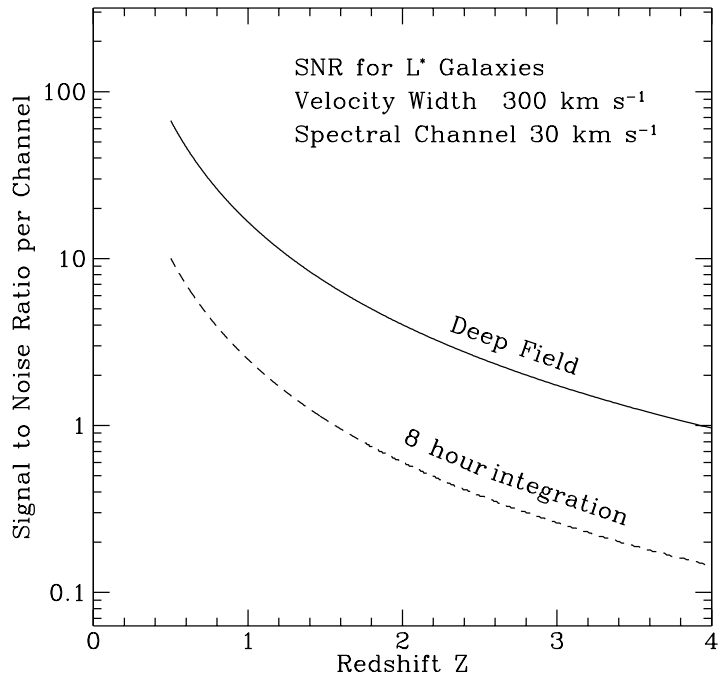}
    \caption{\small Signal to noise ratio for spectral measurements
of an $M_{HI}^*=10^{9.8}$M$_{\odot}$ galaxy as a function of
redshift. S/R ratios are computed for 30~km~s$^{-1}$ channels observing  
galaxies with 300~km~s$^{-1}$ wide emission profiles  
Two cases are considered: {\it lower curve} a 8 hour SKA integration
reaching  $\sigma= 3\mu$Jy and  {\it upper curve} a 360 hour SKA 
Deep Field integration.
}
    \label{snr.fig}
  \end{center}
\end{figure}

\section{Sizes and kinematics of the DLa Clouds}

\subsection{Emission observations}

For nearby galaxies, synthesis mapping techniques provide
measurements of the extent of the HI emission and kinematics
that help to clarify the structure of the galaxies and the
distribution of their dark matter component. These elegant maps and
subsequent analysis typically describe the nearby galaxies with peak
flux densities in the integral profiles of a few 100 mJy with
sensitivity levels around $\sigma \approx 1$~mJy per velocity
channel ($\sim$10 km~s$^{-1}$ wide).

The prospects for obtaining this level of sensitivity 
with $S/N \geq 100$ in high redshift galaxies are not good.
Fig.~\ref{snr.fig} summarizes the expected $S/N$ for 
$M_{HI}^*=10^{9.8}$M$_{\odot}$ galaxies as a function of
redshift. For this example, the galaxies have total
velocity spreads of 300~km~s$^{-1}$, which is observed with
channel spacing of 30 km~s$^{-1}$. A SKA 360 hour
integration cannot achieve $S/N>50$ for redshifts
greater than 0.6 and hits $S/N=10$ for $z=1.25$.

\subsection{Absorption observations}

Fortunately, definitive measurements can be obtained through
high spatial resolution observations of absorption against
extended background radio sources. Fig.~\ref{schematic.fig}
shows that roughly half of the radio sources with redshift
greater than 4 will lie behind DLa absorbing layers.  SKA
sensitivities will permit absorption experiments against
random high redshift radio sources in every field.  In fact,
a standard tool for determining a lower limit to redshifts
for optically blank field radio sources may be to examine the radio
spectrum for narrow absorption lines of redshifted HI.

In fact, considerable progress in assessing the extent and kinematics of
the DLa class of quasar absorption line system can be made
before the commissioning of SKA
with minimal technical adaptation of existing radio facilities.
The technique requires background radio quasars or high redshift
radio galaxies with extended radio continuum emission. Some
effort  needs to be invested in surveys to find redshifted
21cm line absorption against these types of sources.  These
surveys can either key on optical spectroscopy of the quasars to
find DLa systems for subsequent inspection in the 21cm line, or
they can make blind spectral surveys in the 21cm line directly,
once the new wideband spectrometers that are being constructed for
at Westerbork and the new Green Bank Telescope are completed. Then
radio interferometers with suitable angular resolution at the redshifted
21cm line frequency must be used to map the absorption against the
extended background source. This would involve interferometer
baselines of only a few hundred kilometers -- shorter than is
typically associated with VLBI techniques, but longer than
the VLA and GMRT baselines.  The shorter  spacings in the European
VLBI Network and the MERLIN baselines
would form an excellent basis for these
experiments, although considerable effort will be required to
observe at the interference riddled frequencies outside the
protected radio astronomy bands.

\begin{figure}
  \begin{center}
    \includegraphics[width=13cm]{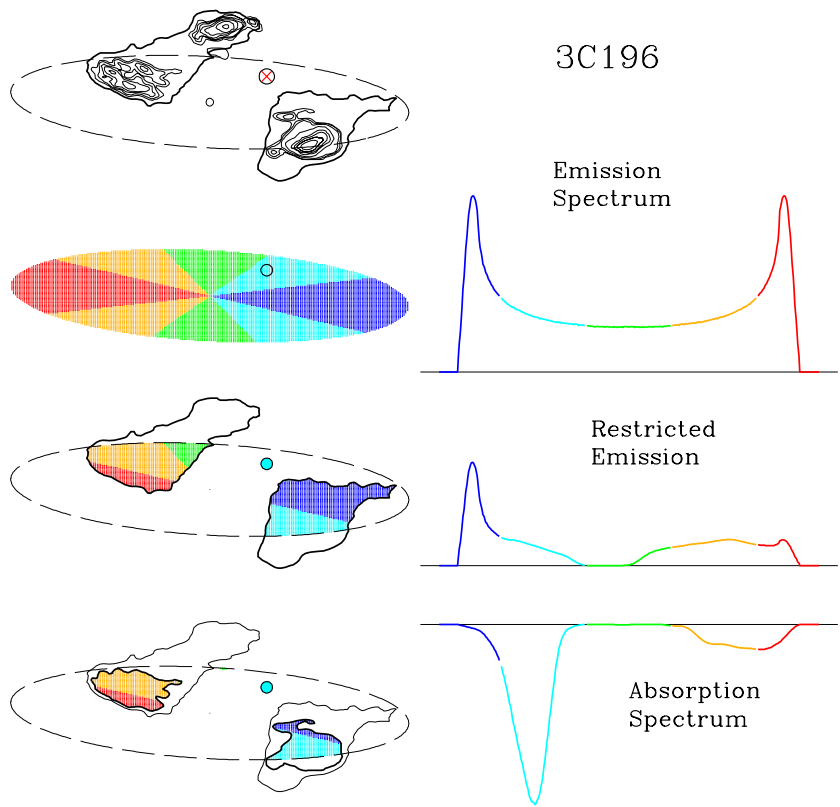}
    \caption{\small Absorption by an intervening disk galaxy 
against an extended background radio quasar. {\it Top:}
Contours of radio continuum emission (Lonsdale 1983 with
the outer radio contour taken from 
the map of Oren as shown by Cohen et al 1996). 
{\it Upper middle:} The velocity field of and emission
profile expected for disk galaxy.
{\it Lower middle:} A spectrum that has been restricted to gas
lying in front of background continuum; in principle, sensitive
mapping could measure the distribution and kinematics for these
clouds in absorption across the face of the radio source.
{\it Bottom:} The integral absorption spectrum obtained by
observing this source with a low angular resolution telescope.
}
    \label{3c196.fig}
  \end{center}
\end{figure}

Fig.~\ref{3c196.fig} shows an example of how these experiments might work.
The top panel shows contours for the radio source 3C196. 
Brown and Mitchell (1983) discovered a 21cm line in absorption at $z=0.437$
against this source in a blind spectral survey. The object has
been the target of intensive optical and UV spectroscopy (summarized by 
Cohen et al 1996), as well as HST imaging to identify the
the intervening galaxy responsible for the absorption (Cohen et al
1996, Ridgway and Stockton 1997). Fig.~\ref{3c196.fig} includes a dashed
ellipse in the top panel to indicate the approximate
extent and orientation of the galaxy identification.

The second panel from the top in Fig.~\ref{3c196.fig} illustrates the 21cm line
emission spectrum typical of nearby HI-rich disk galaxies, observed
by a low resolution (``single-dish'') beam that does not resolve 
the gaseous structure in the galaxy. The rotation of a galaxy
with a flat rotation curve produces the velocity field shown
to the left of the spectrum

For disk systems observed in absorption, the information accessible to
the observer is less, since we can only hope to ever learn about the
gas opacity and kinematics for regions that fall in front of background
continuum. This restricts our knowledge to zones outlined in the
third panel of Fig.~\ref{3c196.fig}. The ``restricted emission'' spectrum is
drawn to illustrate what fraction of the galaxies gas content might
be sensed by a sensitive synthesis mapping observation. A comparison
to the total gas content in the upper spectrum suggests that much of
the important information (velocity spread, for example)
would be measured by a synthesis map of the absorption against
background source.

The single-dish spectrum of the absorption lines observed for an
object like 3C196 is weighted by the regions where the background
continuum has the highest brightness.  As shown in the lower panel,
this weighting emphasizes the bright spots in the radio lobes. 
Clearly sensitive mapping will better recover the information 
lost in the integral spectrum produced by a low angular
resolution observation.  A preliminary look at recent observations
of the $z=0.437$ absorber in 3C196 can be found in de Bruyn et al
(1997).

\section{Direct measurement of the merger rate}

OH megamasers occur in the nuclear regions of merging and heavily 
interacting 
galaxies.  The galaxies are characterized by disturbed morphologies,
strong far infrared emission and heavy extinction at the
center. The brighter OH megamasers can be detected easily at cosmological
distances. Due to the heavy obscuration, these objects are not
especially eye catching, but their strong FIR flux has led to the
identification of a representative sample.

Briggs (1998) argued that the OH megamasers may be a useful tool
in direct measurement of the galaxy merger rate over time,
since these sources would be expected to turn up in radio
spectroscopic surveys, such as a SKA Deep Field.  Simply counting the
number of OH megamasers per volume as a function of redshift would
specify the number of galaxies in the merging phase at that time.
The selection is both immune to obscuration and unbiased with respect
to redshift since the entire radio spectrum (aside from regions of
strong rfi) can be covered with a single telescope.

Fig.~\ref{detecrate.fig} shows estimates for the detection
rate for OH megamaser sources for comparison with the HI emission
from ordinary galaxies. The OH calculations use a constant
comoving density of OH megamaser hosts and the local OH megamaser
luminosity function. For the levels of sensitivities reached by
current radio telescopes (detection levels 0.2 to 1 mJy), the
OH detections should dominate for frequencies below $\sim$1000 MHz.
Given the expectation that mergers were more numerous in the
past, there may be a much high detection rate for OH emission
through the range $1<z<3$ that shown in the figure (Briggs 1998).

\section{Conclusion}

Radio mapping in the redshifted HI line 
with modest spatial resolution radio interferomters
promises to resolve basic questions about how galaxies 
assemble and evolve.  By observing the cool neutral gas that
traces gravitational potential wells of forming galaxies,
the 21cm line provides not only a measure of the neutral
gas content of the Universe over cosmic time scales but also
a method to weigh the dark matter halos.

\section*{Acknowledgements}

The author is grateful to A.G. de Bruyn and J.M. van der Hulst
for valuable discussions.


\section*{References}
\begin{thebibliography}{99}
\bibitem{br1}Briggs, F.H., \& Wolfe, A.M., 1983, ApJ, 268, 76
\bibitem{br2}Briggs, F.H. 1998, A\&A, 336, 815
\bibitem{broe} Broeils, A.H. 1992, A\&A, 256, 19
\bibitem{cal}Calzetti, D., \& Heckman, T.M. 1999, ApJ, 519, 27
\bibitem{coh96} Cohen, R.D., Beaver, E.A., Diplas, A., Junkkarinen, V.T.,
  Barlow, T.A., \& Lyons, R.W. 1996, ApJ, 456, 132            
\bibitem{deb97} de Bruyn, A.G., Briggs, F.H., \& Vermeulen, R.C. 1997,
 http://www.nfra.nl/nfra /newsletter/1997-1/index.htm
\bibitem{gia} Giavalisco, M., Steidel, C.C., \& Macchetto, F.D. 1996a, ApJ,
  470, 189 
\bibitem{gui} Guillemin, P., \& Bergeron, J. 1997, A\&A, 328, 499
\bibitem{paper01}Haehnelt, M.G., Steinmetz, M., Rauch, M. 1998, ApJ, 495, 647
\bibitem{hew}Hewett, P.C., Foltz, C.B., \& Chaffee, F.H. 1993, ApJ, 406, L43
\bibitem{paper02}Jedamzik, K., \& Prochaska, J.X. 1998, MNRAS, 296, 430
\bibitem{paper03}Kauffmann, G. 1996, MNRAS, 281, 475
\bibitem{ken}Kennicutt, R.C., Tamblyn, P., Congdon, C.E. 1994, ApJ, 435, 22
\bibitem{aln0} Lanzetta, K.M., \& Bowen, D.V. 1992, ApJ, 391, 48
\bibitem{lan}Lanzetta, K.L., Wolfe, A.M., \& Turnshek, D.A. 1995, ApJ, 440, 435
\bibitem{LeB} Le Brun, J., Bergeron, J., Boisse, P., \& Christian, C.
 1993, A\&A, 279, 33
\bibitem{ma1}Ma, C.-P., Berschinger, E. 1994, ApJ, 434, L5
\bibitem{ma2}Ma, C.-P., Berschinger, E., Hernquist, L.,
  Weinberg, D.H., \& Katz, N. 1997, ApJ, 440, L1
\bibitem{mad0} Madau, P., Fergueson, H.C., Dickinson, M.E., Giavalisco, M.
  Steidel, C.C., \& Fruchter, A. 1996, MNRAS, 283, 1388
\bibitem{mad}Madau, P. 1998, in The Hubble Deep Field, eds. M. Livio,
  S.M. Fall, \& P. Madau, STScI Symposium Series, astro-ph/9709147
\bibitem{pei}Pei, Y.C., Fall, S.M., \& Hauser, M.G. 1999, ApJ, 522, 604
\bibitem{peti} Petijean, P., \& Bergeron, J. 1994, A\&A, 283, 759
\bibitem{pet} Pettini, M., Smith, L.J., King, D.L., \& Hunstead, R.W.
   1997, ApJ, 486, 665
\bibitem{pet2} Pettini, M., Ellison, S.L., Steidel, C.C., \& Bowen, D.V.
   1999, ApJ, 510, 576
\bibitem{pro}Prochaska, J.X., \& Wolfe, A.M. 1997 ApJ, 487, 73
\bibitem{rid} Ridgway, S.E., \& Stockton, A. 1997, AJ, 114, 511 
\bibitem{rao} Rao, S.M., \& Turnshek, D.A. 1998, ApJ, 500, L115
\bibitem{rob} Roberts, M.S. 1963, ARAA, 1, 149
\bibitem{shcm} Schmidt, M. Schneider, D.P., \& Gunn, J.E. 1994, AJ, 107, 1245
\bibitem{sem} Sembach, K.R., Savage, B.D., Lu, L.,\& Murphy, E.M. 1999, ApJ,
  515, 108
\bibitem{spi99} Spitzak, J.~G., \& Schneider, S.~S. 1999, ApJS, 119, 159 
\bibitem{stav} Staveley-Smith, L. Wilson, W.E. Bird, T.S.,
        Disney, M.J., Ekers, R.D., Freeman, K.C., Haynes, R.F.,
        Sinclair, M.W., Vaile, R.A., Webster, R.L., \& Wright, A.E.
        1996, PASA, 13, 243
\bibitem{ste1} Steidel, C.C. 1990, ApJS, 72, 1
\bibitem{ste2} Steidel, C.C. 1993, in The Environment and Evolution of
  Galaxies, eds. J.M. Shull \& H.A. Thronson, Kluwer Academic Publ., p. 263
\bibitem{ste3} Steidel, CC., Pettini, M., Dickinson, M., \& Persson, S.E.
 1994, AJ, 108, 2046
\bibitem{sted2a} Steidel, C.C., Bowen, D.V., Blades, J.C., \& Dickinson, M.
 1995, ApJ, 440, L45
\bibitem{ste3a} Steidel, C.C., Adelberger, K., Giavalisco, M., Dickinson, M.,
  Pettini, M, \& Kellogg, M. 1998, in The Young Universe, eds. d'Odorico,
  Fontana, and Giallongo, ASP Conference Series. (astro-ph/9804237)
\bibitem{ste4} Steidel, C.C., Adelberger, K.L., Giavalisco, M., \&
  Pettini, M. 1999, ApJ, 519, 1
\bibitem{tay} Taylor, A.R., \& Braun, R. 1999, Science with the 
  Square Kilometer Array: A next generation world radio observatory
\bibitem{tur} Turnshek, D.A.  1997, in Structure and Evolution
  of the Intergalactic Medium from QSO Absorption Line Systems, eds.
  Petitjean, P., \& Charlot, S., p. 263
\bibitem{alb85} van Albada, T.S., Bahcall, J.N., Begeman, K., \& Sancisi, R.
   1985, ApJ, 295, 305
\bibitem{bre} van Breugel, W., de Breuk, C., Stanford, S.A., Stern, D.,
  Rottgering, H., \ Miley, G. 1999, ApJ, 518, L61
\bibitem{wol}Wolfe, A.M.,   Turnshek, D.A., Smith, H.E., Cohen, R.D. 1986,
   ApJS, 61, 249   
\bibitem{w95}Wolfe, A.M., Lanzetta, K.M., Foltz, C.B., \& Chaffee, F.H.
   1995, ApJ, 454, 698
\bibitem{zwa}Zwaan, M., Briggs, F., Sprayberry, D., \& Sorar, E. 1997,
  ApJ, 490, 173
\end{thebibliography}
\end{document}